\documentstyle[aps,prl,preprint,eqsecnum]{revtex}
\newcommand{\beq}{\begin{equation}}

\newcommand{\eeq}{\end{equation}}
\newtheorem{problem}{Problem}[section]
\newcommand{\bp}{\begin{problem}}
\newcommand{\ep}{\end{problem}}
\newcommand{\beqr}{\begin{eqnarray}}
\newcommand{\eeqr}{\end{eqnarray}}

\newcommand{\bma}{\left[ \begin{array}{cccc}}
\newcommand{\ema}{ \end{array} \right] }
\begin{document}
\setcounter{page}{0}
\title
{Thomas Precession, Berry potential and the Meron.}
\author{\large R. Shankar$^{1} $ and  Harsh Mathur$^{2}$}
\address{$^{1}$Departments of Physics and Applied
Physics, P.O. Box 6666,\\
Yale University, New Haven, CT 06511\\
$^{2}$A.T.\&T Bell Laboratories \\ 600 Mountain Avenue
\\Murray Hill NJ 07974 } .\\\date{\today}
\maketitle
\begin{abstract}
We begin with a prior  observation by one of us   that  Thomas
precession in the nonrelativistic limit of the Dirac equation may be
attributed to a nonabelian Berry vector potential.   We  ask what
object produces the nonabelian potential in parameter space, in the
same sense that the abelian vector potential arising in the adiabatic
transport of a nondegenerate level is produced by a monopole,
(centered at the point where the level becomes degenerate with
another), as  shown by Berry.    We find that it is a {\em meron}, an
object in four euclidean dimensions with instanton number ${1 \over
2}$, centered at the point where the doubly degenerate positive and
negative energy levels of the Dirac equation become fourfold
degenerate.

\end{abstract}
\pacs{PACS: 03.65 -w,03.30.+p , 02.40.+m }
\narrowtext
\renewcommand{\theequation}{\arabic{equation}}
%\addtocounter{section}{1}
Consider the adiabatic transport of a nondegenerate level of a
hermitian hamiltonian
around a closed loop in parameter space. If $|\chi (R)\rangle$ is the
instantaneous eigenket of the  hamiltonian at a point $R$ in
parameter space, Berry showed \cite{Berry} that\\
(i) the state vector picks up a phase factor due to a vector
potential
\beq
A_{\mu} = \langle \chi |\partial_{\mu}\chi \rangle
\eeq
in addition to the integral of the energy along the circuit and that
\\
(ii) The source of this potential is a quantized magnetic monopole
located at a point where the level becomes   degenerate with another.
In other words,  the field tensor
$F_{\mu \nu}$ derived from $A_{\mu}$ is that of a monopole.

The proper mathematical framework for describing Berry's work was
established by Simon. \cite{Simon}

It was then pointed out by Wilczek and Zee \cite{WZ} that if the
level being  transported has an  m-fold  degeneracy, the effect of a
closed circuit will now be
given by $U(m)$  matrix obtained by a path-ordered integral of a
nonabelian vector potential. These authors provided many
illustrations.

The references mentioned up to this point, along with many others and
very useful commentary may
be found in the  collection edited by Shapere and Wilczek
\cite{Shapere}.

The problem of two-fold degenerate levels  becoming four-fold
degenerate was discussed in its generality by Avron {\em et al}
\cite{Avron}, using spin systems as a model. The relation of our work
to theirs will be  discussed at  the end.

 Our problem here stems from the observation by one of us
\cite{HM}   that  Thomas precession of electron spin in the
nonrelativistic limit could be understood as arising due to
nonabelian SU(2) Berry potential. The idea is as follows. For the
free Dirac equation, the four levels at each momentum break up into
two (spin) degenerate pairs with equal and opposite energies. Since
there is
a large gap separating the positive and negative levels, they will
not mix under adiabatic evolution. (In practice the coulomb potential
of the nucleus causes the adiabatic evolution). It was then a
straightforward matter to compute
the vector potential
\beq
A_{\mu}^{ab} = \langle  a|\partial_{\mu} b\rangle
\eeq
starting from the degenerate positive energy spinors labeled
by $a$ and $b$ and to show that the effect of this potential was to
produce the Thomas precession.  (It is of course possible to do the
same for the negative energy states. Throughout this paper we will
focus on just the positive energy states. )

Here we ask and answer the following question: {\em What is the
source of this nonabelian potential?} We find it is {\em meron}, an
object with Pontrayagin index equal to ${1 \over 2}$. It is not
necessary to be familiar with merons to follow this paper since their
relevant features  will be described later on.

Let us begin with the Dirac hamiltonian
\beq
H  = \mbox{\boldmath $\gamma $} \cdot {\bf p} + \beta  m
\eeq
where the three $\gamma$'s and $\beta$ are $4 \times 4$ matrices
which anticommute and whose square is unity.  Introduce $4$
{\em euclidean } gamma matrices
\beq
\gamma_{\mu} = (\mbox{\boldmath $\gamma $} , \beta  ) )
\eeq
obeying
\beq
\{ \gamma_{\mu} \gamma_{\nu } \}= 2 \delta_{\mu \nu}
\eeq
 and a four vector
\beq
p_{\mu} = ( {\bf p } , m).
\eeq
{\em Note that the fourth component of $p_{\mu}$ is $m$ and not the
energy.}
Then
\beq
H =  \not{\! p} \equiv \gamma_{\mu} p_{\mu} \ \ \ \ \  (\mu =
1,2,3,0).
\eeq

We shall use the representation
\beq
\gamma_{\mu} = \bma 0 & \sigma_{\mu} \\ \sigma_{\mu}^{\dag} & 0 \ema
\eeq
where
\beq
\sigma_{\mu} = ( \mbox{\boldmath $\sigma $} , iI)
\eeq
and
$\mbox{\boldmath $\sigma $}$ are the Pauli matrices and $I$ is the $2
\times 2$ identity.

Since
\beq
 H^2 = \not{\! p} \not{\! p} = p_{\mu} p_{\mu} \equiv p^2 = {\bf p}^2
+ m^2
\eeq
it follows the eigenvalues of $H$ are $\pm \sqrt{ {\bf p}^2 +m^2 } =
\pm p$. Since there should be a total of  four levels, it follows  they
come in doubly degenerate pairs. The degeneracy may be traced back to
fact that
\beq
H = \gamma_2 \gamma_0 H^{*} \gamma_0 \gamma_2
\eeq
which is Kramers' degeneracy. The fact that the levels come with
equal and opposite energies
is due to the fact that
\beq
\gamma_5 = \gamma_1 \gamma_2 \gamma_3 \gamma_0 = \bma I & 0 \\ 0 & -I
\ema
\eeq
anticommutes with $H$ and is unitary.

Let us now introduce a unitary matrix
\beq
U = {\mbox{\boldmath $\sigma $} \cdot \mbox{\bf p} - im \over p}
\eeq
in terms of which
\beq
H = \bma 0 & p U^{\dag} \\ p U & 0 \ema
\eeq
 and the positive energy eigenvectors are
\beq
|a\rangle = { 1 \over \sqrt{2}} \bma \chi_a \\ U \chi_a \ema \label{fif}
\eeq
where $\chi_a$ and $\chi_b$ are any two orthonormal two-component
spinors. We choose them to be the canonical basis
\beq
\bma 1 \\ 0 \ema  \hspace*{1in} \bma 0 \\ 1 \ema  . \label{six}
\eeq
The nonabelian vector potential is now given by a $2 \times 2$ matrix
\beq
A_{\mu}^{ab} = {1 \over 2}\chi_{a}^{\dag} U^{\dag}\partial_{\mu}
U\chi_b .
\eeq
Since $\chi$ is a canonical basis vector we can write in compact
notation
\beq
A_{\mu} =
{ 1 \over 2} U^{\dag} \partial_{\mu} U.
\label{fubini} \eeq

Note that due to the factor of $1/2$, this is not a pure gauge.

Explicit computation shows that
\beq
A_0 = {-i \over 2p^2} \mbox{\boldmath $\sigma $} \cdot {\bf p} \ \ \
\ \ \ \ \ {\bf A}=  {i \over 2p^2}( m \mbox{\boldmath $\sigma $} +
\mbox{\boldmath $\sigma $} \times {\bf p} )
\eeq

The vector potential has no radial component:
\beq
p_{\mu} A_{\mu} = 0.
\eeq
The next step is to compute the field strength
\beq
F_{\mu \nu} = \partial_{\mu} A_{\nu}  -
\partial_{\nu} A_{\mu}  + [A_{\mu} , A_{\nu} ]
\eeq
Given eq.(\ref{fubini}), it follows that
\beq
F_{\mu \nu} = {1 \over 4} \left[ \partial_{\mu} U^{\dag}
\partial_{\nu} U -  \partial_{\nu} U^{\dag}  \partial_{\mu} U \right]
\eeq

Although  this expression  may be evaluated  in a
 straightforward but tedious manner, it is instructive  to employ   some tricks
analogous to the ones used by Berry in the abelian case.

Let us first introduce some conventions. The kets $|a\rangle,|b\rangle,
|c\rangle$ will
stand for positive energy eigenvectors while $|A\rangle, |B\rangle,|C\rangle$
will
stand for negative energy eigenvectors. We need two lemmas:\\
{\em Lemma 1:}
\beq
\langle \partial a|B\rangle = {\langle  a | \partial H |B\rangle \over E_a -
E_B}
\eeq
 proven by taking the derivative of
$\langle  a|H|B\rangle=0$ and using $\langle \partial a|B\rangle = - \langle  a
| \partial B\rangle$. \\
{\em Lemma 2:}
\beq
\langle  a | \partial H |b\rangle = 0 \ \ \ \ \ \ \ a \ne b
\eeq
proven by taking the derivative of $\langle a|H|b\rangle=0$ for the case $a \ne
b$.

Now we have
\beqr
F_{\mu \nu}^{ab} &=& \partial_{\mu} \langle  a | \partial_{\nu} b\rangle - \ \
(\mu \to \nu ) +
\langle  a | \partial_{\mu}c\rangle\langle  c | \partial_{\nu} b\rangle
 - \ \ \ (\mu \to \nu )\\
&=& \langle \partial_{\mu}  a | \partial_{\nu}b\rangle - \ \ (\mu \to \nu ) -
\langle  \partial_{\mu}a| c\rangle\langle  c | \partial_{\nu} b\rangle
 - \ \ \ (\mu \to \nu )\\
&=&   \langle \partial_{\mu} a|A\rangle\langle A | \partial_{\nu}|b\rangle - \
\ \ (\mu \to \nu
)
\eeqr
where in the last line we note  that the intermediate states are
restricted to those of negative energy. We now invoke Lemma 1 to
write
\beqr
 F_{\mu \nu}^{ab}  &=& {1 \over 4p^2} \langle a|\partial_{\mu}
H|A\rangle\langle A |
\partial_{\nu}H|b\rangle - \ \ \ \mu \to \nu
\eeqr
We now  argue that we can enlarge the set of intermediate states to
include the positive energy states well. First let $a \ne b$. Then no
matter which state $c$ we introduce, it will be unequal to either $a$
or $b$, and the term will vanish by Lemma 2. Finally if $a=b$, then
the contribution from $c$ will vanish if $c\ne a$ thanks to  Lemma 2
and by antisymmetry if $a=b=c$.

We can now use completeness to reach the nice result
\beqr
F_{\mu \nu}^{ab}  &=& {1 \over 4p^2} \langle a|[  \partial_{\mu} H,
\partial_{\nu} H ] |b\rangle\\
&=& {1 \over 2p^2} \langle a|\gamma_{\mu}  \gamma_{\nu}|b\rangle\\
&\equiv & {1 \over 2p^2} \langle a|\gamma_{\mu \nu}  |b\rangle \label{eff}
 \eeqr
where it is understood $\mu \ne \nu$ and where
\beq
\gamma_{\mu \nu} = \bma \sigma_{\mu} \sigma_{\nu}^{\dag} & 0 \\ 0 &
\sigma_{\mu}^{\dag} \sigma_{\nu} \ema
\eeq
 If $\mu $ and $\nu$ are spatial indices, $\gamma_{\mu \nu}$ is
self-dual (same in the upper and lower blocks which are eigenspaces
of $\gamma_5$ with eigenvalue $\pm1$), and anti-self-dual (of
opposite sign in the two blocks) if one of the indices is $0$.

Note however that the $F_{\mu \nu}$ for our problem is not self-dual
or anti-self-dual since it is given by the projection of $\gamma_{\mu
\nu}$ into the positive energy subspace.  Indeed
since $\gamma_5$ anticommutes with $H$, its eigenvectors cannot be
also eigenvectors of $H$.

Using the explicit formulae for the eigenvectors, given by
eqs,(\ref{fif}-\ref{six}) we find that
\beq
F_{\mu \nu} =  {1 \over 4p^2}\left[ \sigma_{\mu} \sigma_{\nu}^{\dag}
+ U^{\dag} \sigma_{\mu}^{\dag} \sigma_{\nu}U \right] .
\eeq

In the problem of Thomas precession, the particle goes around in a
tiny loop (in the nonrelativistic limit) in some plane, say the the
$x-y$ plane, {\em  at fixed $m$}. The relevant field is the
``magnetic field '' along the $z$-axis, given in this limit by
\beq
F_{xy} = B_z = { i \sigma_z \over 2m^2} + {\cal O} ({\bf p})
\eeq
in agreement with Reference 6. (In that paper the vector potential
was given in a different gauge due to a different choice of
eigenvectors.)

But now that we had the field in all of parameter space (and not just
the nonrelativistic region) and we decided to explore some of its
properties. We found that
\beqr
p_{\mu} F_{\mu \nu} &=& 0 = p_{\mu} A_{\mu} \\
 D_{\mu} F_{\mu \nu} & = & \partial_{\mu} F_{\mu \nu} + [A_{\mu} ,
F_{\mu \nu} ] \equiv DF = 0\\
Tr (F\tilde{F} ) &=& -8\pi^2 \delta^{(4)}(p).
\eeqr

The first equations has a counterpart in  Berry's abelian
monopole: there  the magnetic field is radial,  ${\bf R} \times
{\bf B} = 0$, which can be written as $R_i F_{ij}=0$. It is possible
to choose the vector potential  so that it too has  no radial
component. Likewise in our problem, only the circulation in tangent planes is
nonzero.

The second equation tells us that  $F$ solves the Euclidean
Yang-Mills equations. It is however not self-dual or anti-self-dual,
nor is it required to be, since the corresponding action (the
integral of $Tr F^2$) is logarithmically divergent.

The last equation, ( which reflects  the fact that $Tr \ {\bf E}
\cdot {\bf B} = 0$ here)  tells us that the instanton density is zero
everywhere, except possibly at the origin where  the field strength
is singular. Indeed one can argue that there must be a delta-function
singularity there as follows.

The instanton density $Q$ may written as the divergence of a vector
$K_{\mu}$ as follows:
(See for example Reference 7 for details and references)
\beqr
Q &=& -{1 \over 16\pi^2} Tr (F \tilde{F} )\\
 &=& \partial_{\mu} K_{\mu} \label{pont} \\
K_{\mu} &=&  -{1 \over 8\pi^2}\varepsilon_{\mu \nu \alpha \beta} Tr [
A_{\nu} (\partial_{\alpha} A_{\beta} + {2 \over 3} A_{\alpha}
A_{\beta} )]  \label{kay}
\eeqr
Thus the total instanton number inside any volume can be found by
doing the surface integral of $K_{\mu}$. Since $K_{\mu}$ must have
the form $cp_{\mu}/p^4$, by dimensional analysis, we can find $c$ by
considering a point on the m-axis. It is readily seen that $c=
1/4\pi^2$ which fixes
\beq
 Tr (F \tilde{F} ) = -8\pi^2 \delta^{(4)}(p)
\eeq
which in turn implies $Q=1/2$ as per eq.(\ref{pont}).

At this point it became clear that the  field was that of a {\em
meron}. The meron  was first discovered as a solution to Yang-Mills
equations  by De Alfaro {\em et al}\cite{Dealfaro} with instanton
number $1/2$ and invoked by
Callan {\em et al} \cite{CDG} as a configuration that could enter the
functional integral for the Yang-Mills field and explain confinement
by producing the area law decay for the Wilson loop. We paraphrase
their description of the relation between the meron and  the instanton,
which we found very enlightening. We will however do it in terms of
the two-dimensional instanton which arises in the O(3) sigma model
since it is easier to visualize.
Consider  the instanton of unit size which
is obtained by placing a sphere of unit diameter
on
top of the plane and doing a stereographic projection and
assigning to each point on the plane the unit
vector associated with its inverse image:  thus the origin
gets assigned the south pole, the unit vector slowly starts rotating
upwards as we move out and the equator gets mapped into a unit circle .
As we move along this circle, the field lies in the
plane and is radial. As we move further out, the field tilts up even
more and finally at infinity, points upwards. If we measure the
instanton number enclosed within any circle, it will grow from zero
to a half by the time we reach the  unit circle, and reach
unity by the time  we integrate out to infinity. Now we deform the
map as
follows. First we scale down the region inside the unit circle, which
carries the lower hemisphere, to a circle of radius $r<1$; push  the
region outside the unit circle (which carries the upper hemisphere)
to outside a radius $R>1$ and assign  to the entire annulus  $r <1 <
R$, the radial equatorial field.   The instanton density  reaches
the value $1/2$ at $r$, stays fixed out to $R$ (since the entire
annulus maps back to just the equator) and starts growing to unity as
we go past $R$ to infinity. {\em If we now let $r \to 0$ and $R \to
\infty$ we get the meron.} It will look just like the vortex of the
x-y model, have a logarithmically large action, ($\ln R/r$) and carry
no instanton density except at the origin. Our meron is related to
the four dimensional instanton in just the same way: one takes the
instanton, squeezes the region with half the winding
 number into the origin, sends the other half out to infinity and
fills all of space with a configuration that has zero $Tr \ F
\tilde{F}$.

We now relate our work to the general analysis of Avron {\em et al}
of hamiltonians with   Kramers degeneracy,  brought to our notice as
this work was nearing completion. These authors  showed that the
generic hamiltonian with Kramers degeneracy lives in a {\em five}
dimensional Clifford algebra. Let us call the   coefficients $p_{\mu}
= ( {\bf p}, m, M)$. This  corresponds to our adding a term $M
\gamma_5$ to our $H$.  Because of our interest in the physical
problem of the Thomas precession of the Dirac electron, we  sliced
this  space along the plane $M=0$. The relation of our four
dimensional analysis to the five dimensional one is best explained by
a simpler analogy. Consider the monopole that arises in Berry's
treatment of the abelian problem. It lives in three dimensions. It
can be  described by a vector potential  with no radial component and
a field tensor which too has circulation only in the  planes
tangential to the position vector.  By surrounding the monopole  with
a unit sphere $S^2$ and measuring the flux (in proper units) we can
obtain a topological invariant of the map (first Chern number) from
the space of states to $S^2$.  On this sphere the field strength is
uniform and finite. Suppose one were now to slice this space along
the plane $z=0$. On this plane one will see a radial field of the
x-y vortex, i.e., the meron. The unit sphere $S^2$  will be sliced
along the unit circle $S^1$ enclosing the meron. Consider now the
integral of the monopole vector potential $A_{\mu}= \langle \chi
|\partial_{\mu} \chi\rangle$ along this circle.
By Stokes' theorem this equals the flux enclosed in {\em any} surface
bounded by  it. If one uses as the surface the upper hemisphere of
the $S^2$, one gets half the topological   charge, the contribution
being uniform on the hemisphere. The other choice for the surface is
the interior of the circle in the x-y plane, i.e., the disc with
$S^1$ as its boundary.  In the latter case  the flux is zero except at the
origin
where the contribution is $1/2$ due to all the flux flowing upwards into
the upper hemisphere. {\em Thus the person restricting
himself to the plane can reconstruct the topological index associated
with the $S^2$ by doing an integral on the $S^1$ he sees, thanks to
Stokes theorem.}  (He must then remember to double to the answer for
the lower hemisphere in this symmetric problem, or more generally,
take the difference with another oppositely
oriented line integral of the vector potential that describes the
lower hemisphere.)

Returning to  our problem, we must begin with  $H = p_{\mu}
\gamma_{\mu}= \not{\! p}$ in five dimensions, find its eigenvectors
and define $A_{\mu} = \langle \chi | \partial _{\mu} \chi \rangle$ and find the
ten component field strength ${\cal F}_{\mu \nu}$. The result will be
just as in eq. (\ref{eff}) (thanks to the lemmas)
\beq
{\cal F}_{\mu \nu} = {1 \over 2 p^2} \langle a| \gamma_{\mu} \gamma_{\nu}
|b\rangle
\eeq
 with the obvious change in the definition of  $p$ and
the range of indices. The field is still singular at
the origin in five dimensions, obeys $p_{\mu}{\cal F}_{\mu \nu}=0$
and D${\cal {F}} =0$. The former implies that the   field is
tangential. (This is true in any   gauge. Also one can pick a
gauge,as we did,
in which $A_{\mu}$ itself has no radial component.)  Thus at each point
$p_{\mu}$,
${\cal F}$ has only six nonzero components with ``transverse''
indices. We shall refer to them  as $F$.  {\em  This configuration is thus
a natural generalization  of the Berry monopole configuration in three
dimensions. }

Avron {\em et al} show that
if we surround the origin by a unit sphere $S^4$,
\begin{itemize}
\item The (suitably normalized) integral of $Tr\ F \tilde{F}$ on the
sphere  (where  the dual is with respect to the epsilon symbol
$\hat{p}_{\lambda}\varepsilon_{\lambda \mu \nu  \alpha \beta}$)
is $\pm 1$ and measures a topological invariant (second Chern number
or Pontrayagin index)
of the map from space of states to $S^4$.
\item  $F$  is
  is self -dual or anti-self-dual depending on the sign of the energy.
\end{itemize}
Since all points on the sphere are equal by symmetry, let us go to
one with just $M = 1$, rest equal to zero. Now
$\hat{p}_{\lambda}\varepsilon_{\lambda
\mu \nu  \alpha \beta}$  becomes just $\varepsilon_{ \mu \nu  \alpha
\beta}$, the six nonzero components of ${\cal{F}}$ have their
indices going from $0$ to $3$.  Since $H=\mu \gamma_5$ here, the
eigenstates of $H$ are indeed eigenstates of $\gamma_5$ and the field
is self(anti)-self-dual as observed earlier with respect to
$\varepsilon_{ \mu \nu  \alpha \beta} $.

By  focusing on the Thomas precession of the physical Dirac equation,
we sliced
this space along $M=0$. Our plane contains the singularity at the
origin and the sphere $S^3$ which is the equator of the $S^4$. The
integral on this sphere of $K_{\mu}$, defined in  eq.(\ref{kay}),
gives, by Stokes
theorem, either  half the topological charge associated with the
$ S^4$ (if we view the $S^3$ as the boundary of the hemisphere of
$S^4$) or the meron charge if we view it as the boundary of the disc
(or ball)
contained in the slice.
(In the language of forms, since $Tr \ F \tilde{F}$ is closed on the
$S^4$, it can be   can be written locally as the derivative of a
three form, which is just the dual of $K_{\mu}$.  The integral of the
topological density associated with the upper hemisphere can be
written via Stokes' theorem as the  integral over  the equatorial
$S^3$ of this three form. But by Stokes theorem, this also equals the
integral of $Tr \ F \tilde{F}$ within the ball in the $M=0$ slice
bounded by the same $S^3$, which gives now the meron's winding number
of a half.)

To summarize, we looked at the adiabatic evolution of the Dirac
hamiltonian and asked what singularity in parameter space produces
the nonabelian vector potential which is behind the Thomas precession
and found that it was the meron. Although we computed the
connection for all possible transports in parameter space, {\em only
closed paths in momentum at fixed $m$ arise in atomic physics} since
$m$ is invariant. On the other hand, in cosmological  models in which
$m$ can vary (say with some Higgs field, which gives the fermion its
mass) other components of the field can produce observable effects.
The connection to merons is amusing and possibly  has other
implications.
Callan {\em et al}  offer many arguments for the meron's previleged role
before proposing it as a configuration that can
 make important
contributions to
the Yang-Mills functional integral. That the  meron  appears naturally in the
Berry
analysis provides yet another argument.
As for the {\em topological}
ideas discussed here, they are naturally
subsumed by  the very general five dimensional analysis of
Avron {\em et al} .  However the four dimensional slice produces singular
configurations interesting in their own right,
 just like the x-y model vortices that arise on slicing the monople.
We hope  readers will find the concrete example of
the Dirac problem (with its relation to Thomas precession) and the
relation between the $S^4$ instanton of Avron {\em et al} in five
dimensions and the meron that appears in our four dimensional  slice,
a useful addition to our knowledge of instantons, merons, Thomas
precession and the Berry connection.

RS  acknowledges some wonderful discussions with Greg Moore. This
report was supported by an NSF Grant
DMR 9120525.

\end{document}